# 3D elastoplastic simulation of ski-triggered snow slab avalanches

Marcus Landschulze[1]

**Abstract**

The stability of dry-snow avalanches is strongly dependent on the interaction between the snow slab above a weak-layer and, as presented in this work, the skier induced load. This induced load causes an additional stress field on the slab which eventually triggers an avalanche. I present the results of 3D finite element simulations in an elastoplastic domain. The plastic deformation of the weak-layer follows the Mohr-Coulomb-Cap model which provides a more realistic model as a pure elastic approach. I investigate how the stress field on top of the weak-layer changes if one is skiing down-slope or parallel to the slope. A layered snow slab changes the stress on top of the weak-layer and to investigate these changes I simulated two different representative layered slabs. One containing only soft layers to investigate how the weak layer is affected by the ski induced stress and the other hard-soft-hard layer to examine bridge effects caused by the hard layers. A hard layer in the snow slab forms a sort of bridge which spreads the induced stress over a larger lateral distance, at the same time decreasing the stress to the layers below the "bridge". Furthermore, I show a possible connection between the plastic deformation and the critical crack length.

keywords: elastoplastic simulation, ski triggered avalanches, 3D simulation, snow stability simulation

## 1 Introduction

Human and naturally triggered snow avalanches occur throughout the winter season in mountainous regions around the world. In the last decade a significant increase in skiing activities in the backcountry has resulted in an increase in avalanche incidents and fatalities. This season (2018/2019) has already seen a ten-fold increase in fatalities by avalanches and hundreds more injured and affected.

A slab avalanche is often triggered by a skier; if the additional mechanical load of the skier induces a stress field in the weak-layer, the weak-layer fabric will collapse (e.g. Heierli et al. 2011).

In order to evaluate snow slope instability simplified assumptions are used to account for the balance between the shear strength in the weak-layer and the shear stress caused by the stress induced skier load (Monti et al., 2016; Schweizer et al., 2006). The first attempt to solve skier induced trigger of an avalanche was done by Föhn (1987) where the author introduced a skier stability index which modeled the skier as an static load on an elastic and isotropic snow slab. Jamieson and Johnston (1998) modified and updated the skier stability index to include skier penetration into the snow slab. However, snow slab avalanches are considered as a sequence of failure initiation and crack propagation (Gauthier and Jamieson, 2006; Schweizer et al., 2003). The elastic slab properties and the strength of the weak-layer are considered as important properties and the interaction between the weak-layer and the slab is crucial for the avalanche release (Herwijnen and Jamieson, 2007; McClung and Schweizer, 1999).

---

[1] Western Norway University of Applied Sciences, Department of Computer science, Electrical engineering and Mathematical sciences, P.B. 7803, 5020 Bergen, Norway E-mail: marcus.landschulze@hvl.no

In order to improve the evaluation of the snow instability the failure initiation and crack propagation needed to be considered in a new stability criterion. Reuter et al. (2015) showed that the combination of failure initiation and crack propagation in the weak-layer improve significantly the snow instability estimation. But this approach does not include the slab load on the weak-layer which was added recently by Gaume and Reuter (2017).

In this paper I present a three-dimensional slab-weak layer system using two simplified snow layer profiles similar to Habermann et al. (2008). These profiles are based on finite element simulations to compute the 3D stress filed and how the slab and skier induced stress field propagates inside the snow slab. The simulation results are focusing on the stress field and volumetric plastic strain on top of the weak-layer, but bridging effects caused by hard snow slab layers are investigated as well. In order to simulate plastic behavior volumetric plastic strain was calculated by using the Mohr-Coulomb-Cap (MCC) model because Reiweger et al. (2015) show a good agreement between the MCC model and measured field data. I show further how the snow layers characteristics change the stress propagation and how plastic deformation is connected to the critical crack length.

## 2 Methods

Finite element simulation methods are an effective approach for forward modelling and in this study I present the result as applied to a three-dimensional slab-weak layer system. The simulation volume considers isotropic solid mechanics with linear elastic materials and an elastoplastic weak-layer. The simulation was performed with the commercial tool Multiphysics. The mathematical model presented in this work essentially corresponds to the workflow in Multiphysics.

*Elastoplastic behavior*

The elastoplastic model is described through linear isotropic Young's modulus $E$, Poisson ratio $v$ and density $\rho$. In this simulation I selected a stationary solver which simplifies the equation of motion for a linear elastic material to:

$$0 = \rho \frac{\partial^2 u}{\partial t^2} = \nabla FS + F_v , \qquad (1)$$

with $F$ the deformation gradient and $S$ the Second-Piola-Kirchhoff stress tensor. $F_v$ is the volume force applied to the model. Due to assumed small strains the values of $S$ are the same as the Cauchy stress tensor $s$ and equation 1 is simplified to:

$$0 = \rho \frac{\partial^2 u}{\partial t^2} = f_v - \nabla s, \qquad (2)$$

with $f_v$ the volume force vector. This simplification computes the pressure $p$ which corresponds to the volumetric part of the Cauchy stress.

$$p = \frac{-1}{3} trace(s) \qquad (3)$$

The deviatoric part can be easily calculated as:

$$s_d = s + pI, \qquad (4)$$

with $I$ as the unity matrix. The second invariant of the deviatoric stress $J_2$ will be later used for the yield function and can be expressed as:

$$J_2 = \frac{1}{2} s_d : s_d = \frac{1}{6}((\sigma_{11} - \sigma_{22})^2 + (\sigma_{22} - \sigma_{33})^2 + (\sigma_{33} - \sigma_{11})^2 + \sigma_{12}^2 + \sigma_{23}^2 + \sigma_{13}^2), \qquad (5)$$

where the ":"-symbol indicates a contraction over the indices.

*Weak layer failure criterion (effective plastic strain)*

Weak snow layers have a distinct elastic regime, where the deformations are recoverable. But when the stress level exceeds the yield point permanent plastic strain will appear. The yield criterion defines at which stress plastic deformation occurs. Stresses below the yield point are purely recoverable deformations, but above the yield point recoverable and permanent types of deformations are produced. In this simulation the Mohr-Coulomb-Cap (MCC) model was used (e.g. Reiweger et al. 2015). The MCC can causes numerical difficulties at the corners of the yield surface and therefore the Drucker-Prager approach was used to smooth the MCC. The yield function is defined as follows:

$$F_y = \sqrt{J_2} + \alpha I_1 - k = 0 \qquad (6)$$

Where α and k are the Drucker-Prager coefficients, $I_1$=trace($\boldsymbol{\sigma}$)=3$p$ the invariants of the pressure *p*. The relation between the Mohr-Coulomb coefficients and Drucker-Prager can be expressed as:

$$\alpha = \frac{2}{\sqrt{2}} \frac{\sin\phi}{3 \pm \sin\phi} \quad \text{and} \quad k = \frac{2\sqrt{3} c \cos\phi}{3 \pm \sin\phi}, \qquad (7)$$

where φ = angle of internal friction and c = cohesion.
In order to have a more realistic yield criterium I used an elliptic cap to overcome the problem that snow cannot bear infinite loads. The cap surface was used to limit the stress $\sigma_{fast}$ and $\sigma_{slow}$ and calculated as the cap-surface (e.g. Lee and Huang 2015):

$$R = \frac{\sigma_{slow} - \sigma_{fast}}{c + \sigma_{fast} \tan\phi} \qquad (8)$$

*Critical crack length*
The critical crack length $a_c$ characterizes the crack propagation in the weak layer caused by the area where the skier-induced stress exceeds the shear stress of the weak-layer. I follow the formulation proposed by (Gaume et al., 2017):

$$a_c = \Lambda \left[ \frac{-\tau + \sqrt{\tau^2 + 2\sigma(\tau_p - \tau)}}{\sigma} \right], \qquad (9)$$

with σ=ρ g D cos ψ, D the slab thickness, ψ the slope angle, g = gravitation constant, τ = gravitational force, $\tau_p$ = τ - Δτ, Δτ = additional ski induced shear stress and the characteristic length Λ:

$$\Lambda = \sqrt{\frac{E' D D_{wl}}{G_{wl}}}, \qquad (10)$$

where E' = E/(1- $v^2$).

The combination of the failure initiation and crack propagation on the stability index can be calculated as follows (e.g. Gaume and Reuter 2017):

$$S_p = \frac{a_c}{l_{sk}}, \qquad (11)$$

with $l_{sk}$ as skier crack length. Later I will use the volumetric plastic strain area to get the $l_{sk}$.

## 3 Model description
The 3D geometry and layer structure is based on the 2D finite element simulation described in (Habermann et al., 2008). The inferred simplified model geometry is shown in Figure 1 and for simplicity all layers are assumed to be horizontal and parallel to each other. However, these simplifications are a good approximation to real snowpack conditions considering the small size of the model volume.

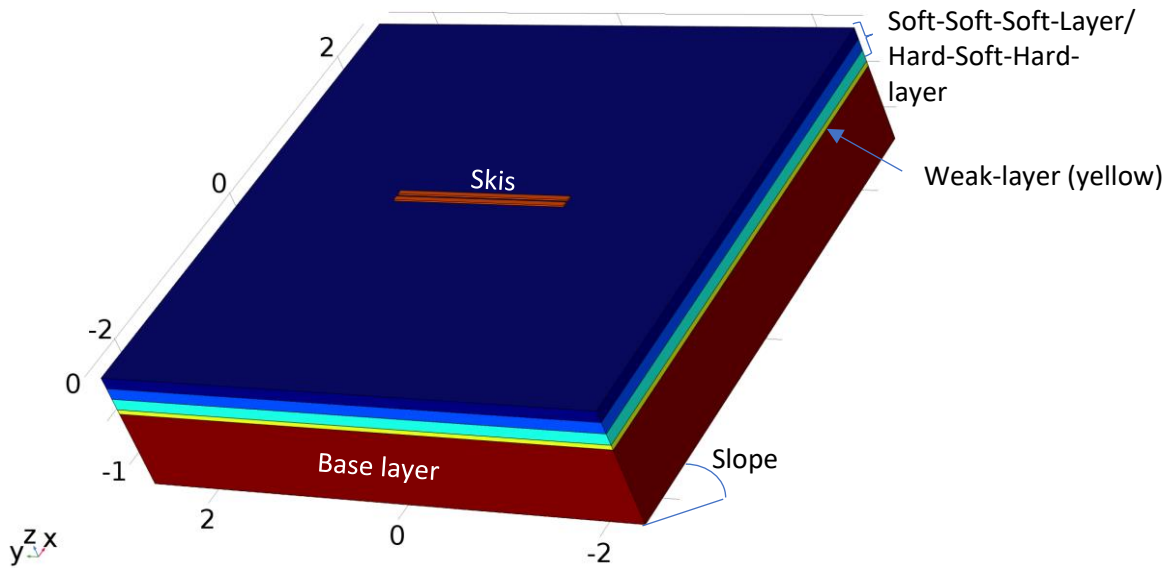

**Figure 1** Schematics of the snow layer structure. The top three layers are 5m x 5m and 0.12m thick, the weak-layer is 0.05m thick and the base-layer 1.21m. The skis are 1.75m long, 0.05m wide and 0.015m thick. The distance between the skis is 0.08m.

*Geometry and Mesh*

Meshing is aimed at minimizing the number of elements and hence computing time, while simultaneously providing sufficient coverage in areas of rapid changes. This is achieved by high density of small elements in areas of expected large changes, i.e. heterogeneities and edges, while fewer, larger elements sufficiently represent the homogeneous areas (Figure 2). The simulations were run on a 16-core cluster computer with 128 GByte RAM. An average number of 901083 degrees of freedom (DOF) were solved and the computing time was 10 minutes on average.

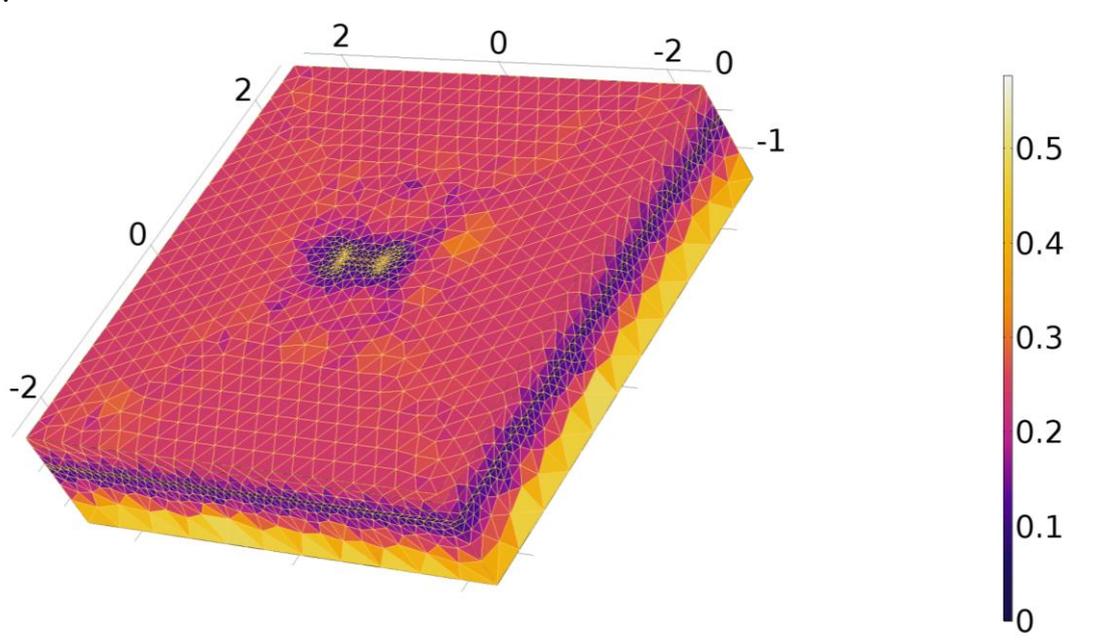

*Figure 2* Dynamic mesh of the 3D model. In the area of interest the mesh-size is denser compared to e.g. the base. The color-code indicates the mesh-size in meter.

*Boundary conditions, initial conditions and mechanical Load*

The boundary on the bottom of the model contains a fixed constraint which forbids all displacement in the vertical direction below the base layer. All other boundaries are considered as free boundaries which means that displacement is possible in these directions. This fixed constraint was selected to simulate a rock layer and the mechanical load used will be not able to deform or displace the layer below the base. In this simulation I assume that the snow layers are in equilibrium hence the gravitational force of the slab weight was not considered during the simulation (g=0). For the correct values after the simulation the following equation will be added to all necessary stresses:

$$\tau_g = -g\rho Z sin\psi. \tag{12}$$

Furthermore, no initial displacement fields or structural velocity fields are used. The only source is the induced skier weight which is split between the two ski surfaces as a constant load weight.

*Hardening rules*

The simulation in this work assumes that the hardening and softening depends on the volumetric plastic deformation and when the volumetric plastic strain increases the weak-snow layer may fracture under tension.

## 4 Results

In order to test the model and the correct setup for the simulation, a benchmark against an established simulation model is needed. Using the package Multiphysics, I performed a 2D simulation equal to the setup described in (Gaume and Reuter, 2017). Figure 3 confirms a similar result as described in Figure 8 in the reference paper.

After confirming that the software tool did produce similar results as reported, I constructed a 3D model to simulate the stress- and shear-stress field inside the model volume.

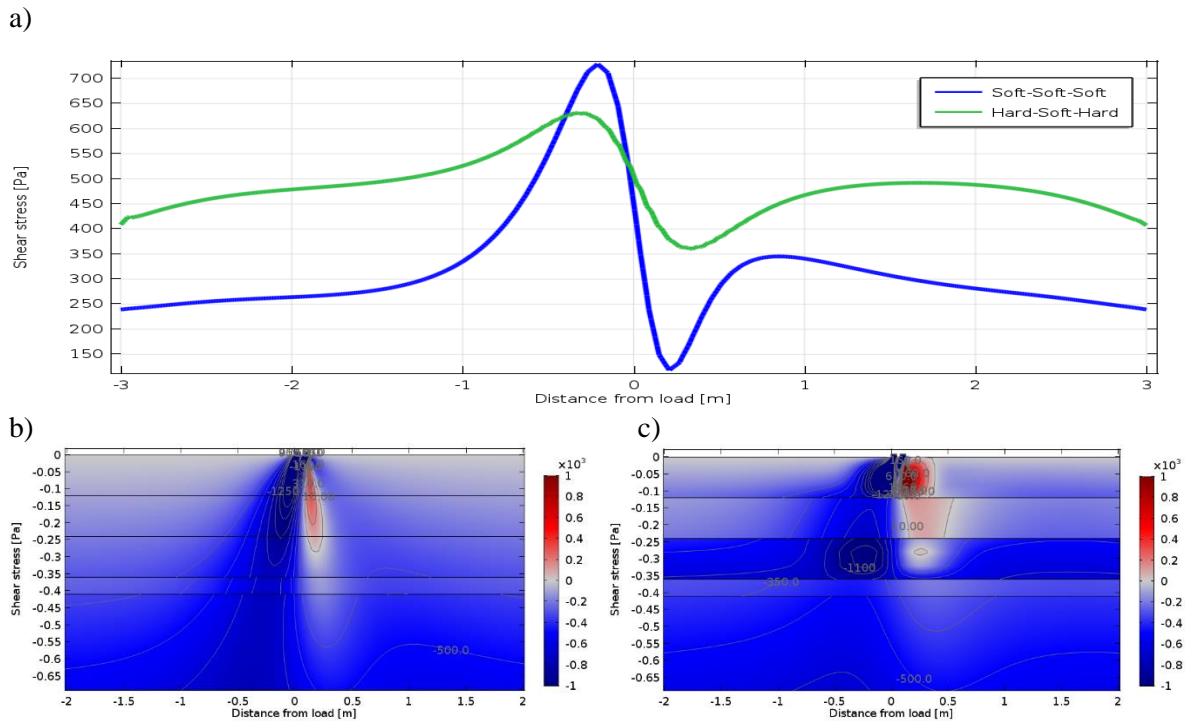

**Figure 3** *Benchmark simulation result similar to that described in* (Gaume and Reuter, 2017). *a) shows two shear-stress curves at the depth of the weak-layer. Zero distance from load is in the center between both skis. In blue Soft-Soft-Soft layer characteristic (F-F-F hand hardness index) and in green Hard-Soft-Hard (1F-F-1F). b) illustrates in a 2D cross-section below the ski for F-F-F and c) 2D cross-section below the ski 1F-1F-1F*

In this work I show the results of two different layer characteristics, following (Habermann et al., 2008), hardness-profile *a* and *d* with a soft-base. Other hardness-profiles are tested as well, but not shown in this work. Table 1 in the appendix A shows the used model parameters for further details. The following results include the shear stress due to the slab weight accordingly to the depth, if not otherwise stated.

Interpreting 3D slices provides a good overview of the model volume, but difficult to evaluate. Therefore, I show only the 3D normal stress $\sigma_n$ as an example shown in figure 4. I will focus on 2D slices, but the simulation was always performed in 3D.

*Normal stress*

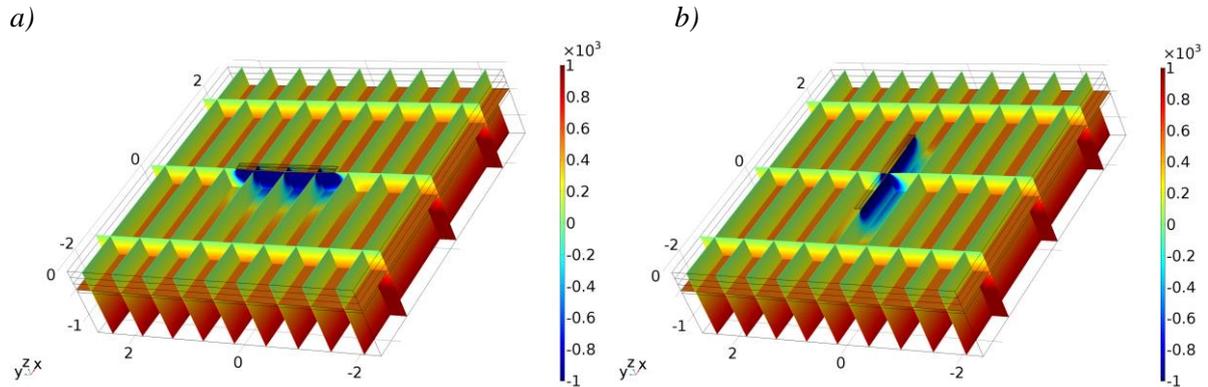

***Figure 4*** *3D normal stress $\sigma_n$ a) with ski parallel to the slope (38 degree). b) with ski down-slope for Soft-Soft-Soft layer characteristic (F-F-F hand hardness index). Color-code is in Pa and distance in meter.*

*2D slices from the 3D model volume*
In this section I selected two vertical slices, one with Soft-Soft-Soft layer characteristics where the snow slab is separated into three similar layers which provides a constant density and Youngs modulus over the whole slab with thickness D of 0.36m and one with Hard-Soft-Hard layer characteristics (similar to Habermann et al. 2008) to show as an example how the stress field propagates through the snow pack under different conditions.

Figure 5a illustrates the normal stress $\sigma_n$ when the skis point down-slope with the model containing Soft-Soft-Soft layer characteristics. The normal stress immediately below the skis is -2330 Pa and reaches -414 Pa on top of the weak-layer. The ski induced stress field is pointing slightly in the slope direction (to the left), but the major stress direction is still pointing downwards. The skis are also bending, and the deepest bending point is slightly moved in the direction of the slope. The ski response is not part of this work and will not be described further.
In Figure 5b the ski direction is like that shown in Figure 5a - pointing down-slope - but the model has Hard-Soft-Hard layer characteristics. The normal stress immediately below the skis is -1580 Pa, this being 750 Pa smaller in normal stress compared to the soft layer in Figure 5a. On top of the weak-layer the normal stress is -108 Pa which is 306 Pa less normal stress compared to the same weak-layer in 5a. The normal stress field passes the soft layer (layer 2 from top) with reduced stress and is more focused compared to the following hard layer (layer 3). In this second hard layer (layer 3) the stress field dissipates and decays much faster over a greater area. The ski induced normal stress field points slightly in the direction of the slope (to the left), but with the major stress pointing downward.
Changes in the normal stress field compared to the slab-load stress can be seen in the entire simulation volume.

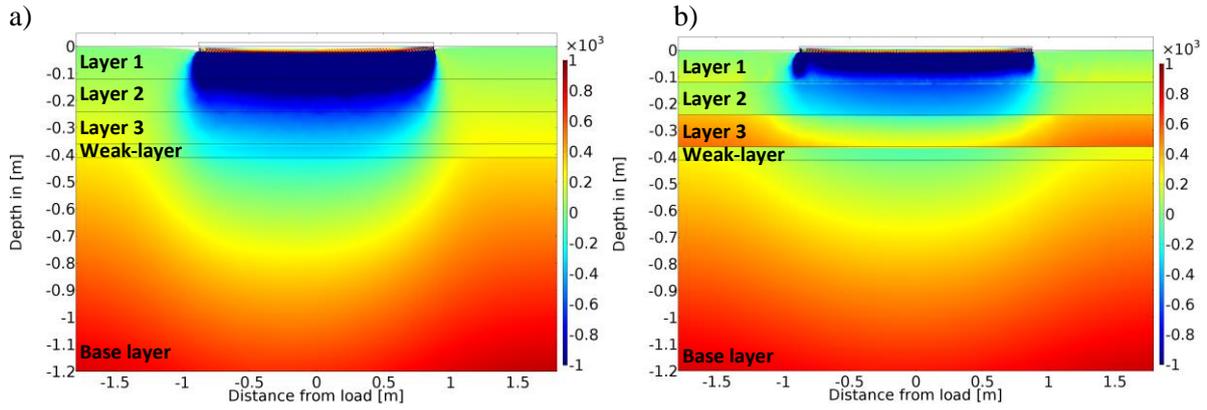

**Figure 5** normal stress $\sigma_n$ with ski down-slope a) for Soft-Soft-Soft layer characteristic (F-F-F hand hardness index) and b) for Hard-Soft-Hard (1F-F-1F hand hardness index). The color-code is in Pa and the deformation is 10 times exaggerated.

Figure 6a is similar to figure 5a, but here the skis are parallel to the slope. The normal stress below the ski is -1574 Pa and decays down to -91 Pa on top of the weak-layer. The stress field is pointing slightly in the direction of the slope (to the left). Figure 6b shows the stress field parallel to the slope with -1580 Pa immediately below the skis. The first hard layer comprises a focus stress field pointing downward, this stress field unravels in the soft layer and decays rapidly at the last hard layer. On top of the weak layer the stress field is decreased to -95 Pa.

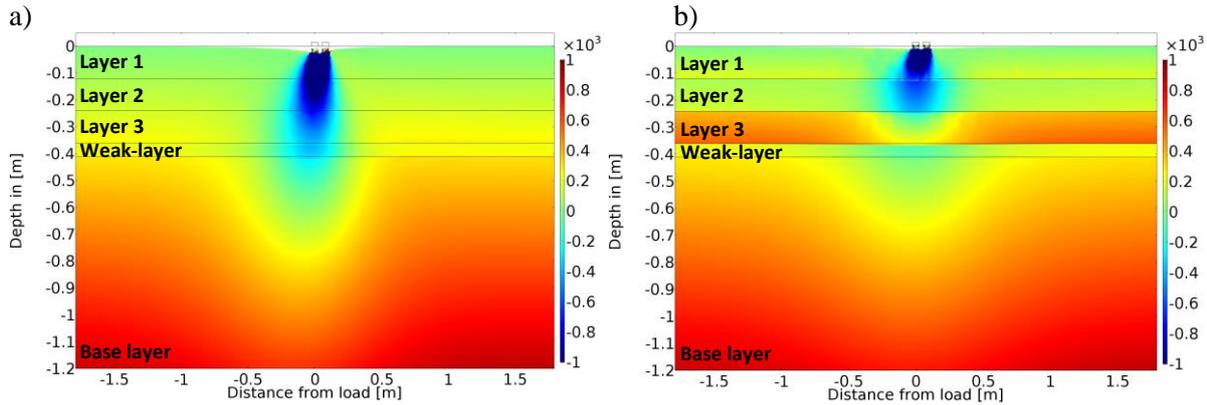

**Figure 6** normal stress $\sigma_n$ with ski parallel to the slope a) for Soft-Soft-Soft layer characteristic (F-F-F hand hardness index) and b) for Hard-Soft-Hard (1F-F-1F hand hardness index). The color-code is in Pa and the deformation is 10 times exaggerated.

*Weak-layer*
In order to better understand the weak-layer response, three different weak-layer characteristics are investigated: Stress, volumetric plastic strain and critical crack length. The following figures show a horizontal slice on top of the weak-layer. All horizontal slices are extracted out of the 3D simulation results. I will emphasis in this work the normal stress $\sigma_n$ as the compressional stress, and the two shear stresses $\tau_{xz}$, $\tau_{yz}$. If not other stated all shear stresses include the slab load (equation 12).

*Skis parallel to the slope*
The following figures 7 to 9 show the stress response where the skis are parallel to the slope. The center of the skis is 0 m and are orientated in the y-direction (parallel to length).
Figure 7 illustrate the normal stress $\sigma_n$ of both layer characteristics. Figure 7a shows a high compressional stress on the left side of the skis with a maximum compressional stress of -414 Pa. This compressional stress field is asymmetric towards the slope.
The compressional stress field in Figure 7b is more symmetrical compared to Figure 7a, but still on the left side of the skis. The maximum value of -108 Pa is located in the center of the stress field.

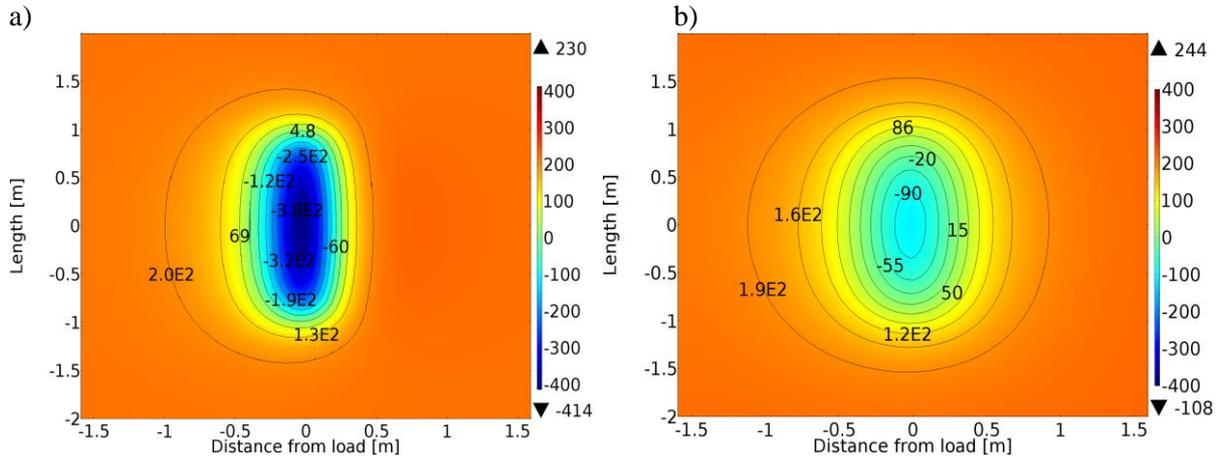

*Figure 7* normal stress with skis parallel to the slope a) $\sigma_n$ Soft-Soft-Soft layer characteristic (F-F-F hand hardness index). b) $\sigma_n$ Hard-Soft-Hard (1F-F-1F hand hardness index). Color-code is in Pa and the values at the black triangles are max-min.

Figure 8a shows the $\tau_{xz}$ shear stress (F-F-F) with a positive value of 290 Pa on the right side of the skis and nearly 0 Pa on the left side of the skis in the direction of the slope. Similar behavior is shown in figure 8b (1F-F-1FH) but less focused and the stress value left of the skis is approx. half of the value on the right side.

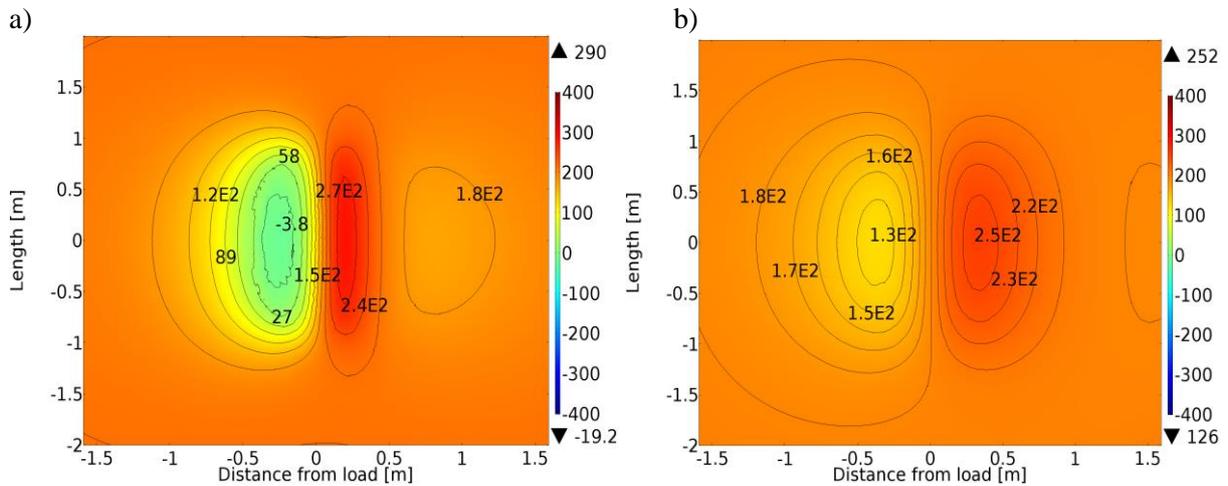

*Figure 8* shear-stress with skis parallel to the slope a) $\tau_{xz}$ Soft-Soft-Soft layer characteristic (F-F-F hand hardness index). b) $\tau_{xz}$ Hard-Soft-Hard (1F-F-1F hand hardness index). Color-code is in Pa and the values at the black triangles are max-min.

The shear stress $\tau_{yz}$ in figure 9a is focused at the start and end of the skis. Both focused stress areas have positive values and are located on the left side of the skis in the direction of the slope. A similar response is shown in figure 9b but is less focused and with less stress values.

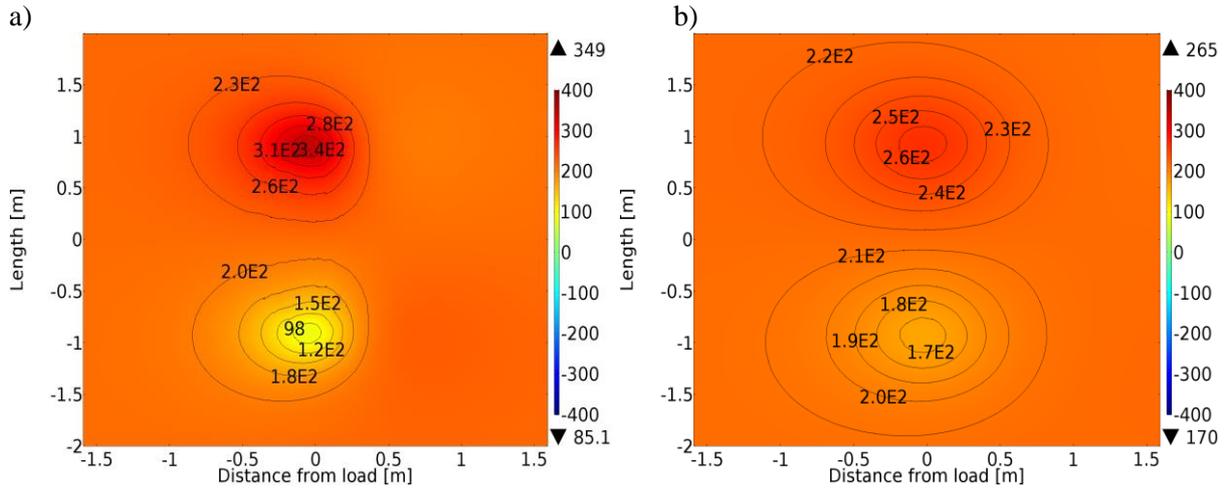

*Figure 9* shear-stress with skis parallel to the slope a) $\tau_{yz}$ Soft-Soft-Soft layer characteristic (F-F-F hand hardness index). b) $\tau_{yz}$ Hard-Soft-Hard (1F-F-1F hand hardness index). Color-code is in Pa and the values at the black triangles are max-min.

The horizontal stress field $\tau_{xy}$ in figure 10 is free of the slab weight but provides high stress amplitudes compared to the other stresses. Figure 10a shows two focused stress areas at the start and end of the skis on the left side. At 1 meter distance from the skis a second focus area is visible with less amplitude. The areas are symmetrical, but different in amplitude. Figure 10b is like Figure 10a but is less focused and with less stress values.

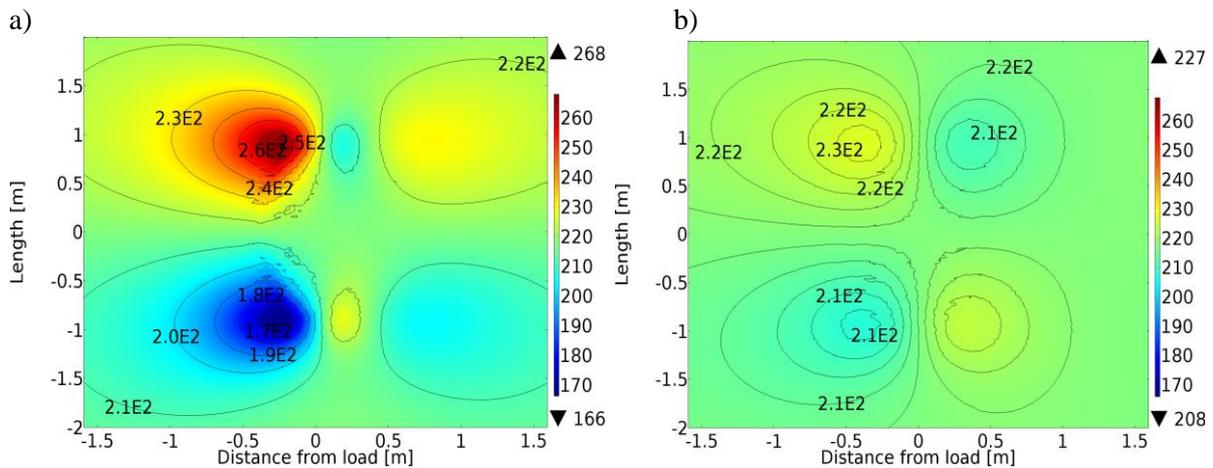

*Figure 10* shear-stress with skis parallel to the slope without slab weight added *a)* $\tau_{xy}$ Soft-Soft-Soft layer characteristic (F-F-F hand hardness index). b) $\tau_{xy}$ Hard-Soft-Hard (1F-F-1F hand hardness index). Color-code is in Pa and the values at the black triangles are max-min.

*Ski down-slope*
The following figures 11 to 14 show the stress fields where the skis are pointing down-slope. The center of the skis is at 0 m and are orientated parallel to distance from the load (horizontal). The slope direction is to the left.
Figure 11 shows the normal compressional stress of the two layer characteristics used. The stress field focus shown in figure 11a is focused around the skis and displaced in the direction of the slope to the left. The Figure 11b is similar to Figure 11a but is less focused but with similar maximum values.

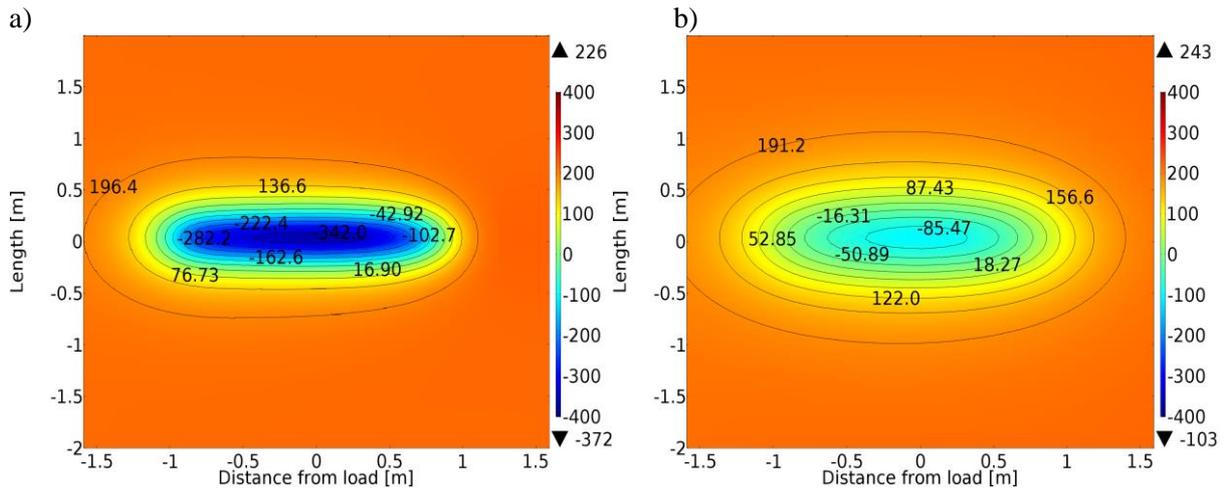

*Figure 11 normal stress with skis down-slope. a) $\sigma_n$ Soft-Soft-Soft layer characteristic (F-F-F hand hardness index). b) $\sigma_n$ Hard-Soft-Hard (1F-F-1F hand hardness index). Color-code is in Pa and the values at the black triangles are max-min.*

The shear stress $\tau_{xz}$ in Figure 12a is more focused at the start and end of the skis compared to figure 11 with nearly 0 Pa at the start point of the skis and 229 Pa at the end of the skis. The stress field is displaced towards the slope. Figure 12b shows a similar response but is less focused with similar stress values.

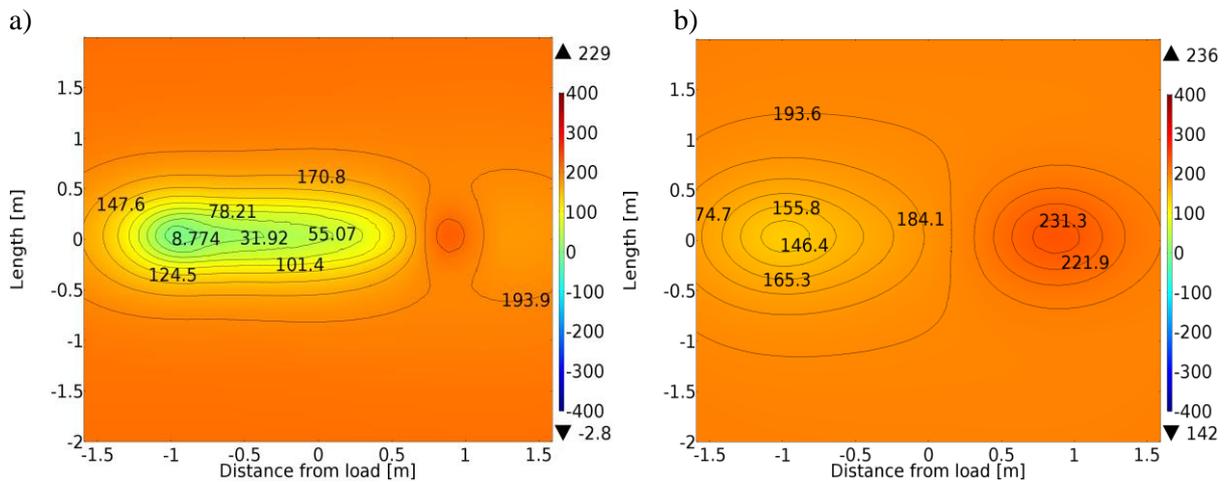

*Figure 12 shear stress with skis down-slope. a) $\tau_{xz}$ Soft-Soft-Soft layer characteristic (F-F-F hand hardness index). b) $\tau_{xz}$ Hard-Soft-Hard (1F-F-1F hand hardness index). Color-code is in Pa and the values at the black triangles are max-min.*

The shear stress field $\tau_{yz}$ shown in Figure 13a is focused around the skis but displaced towards the slope. The stress values are higher compared to $\tau_{xz}$ in Figure 12. Figure 13b is less focused and displaced but with similar stress field values.

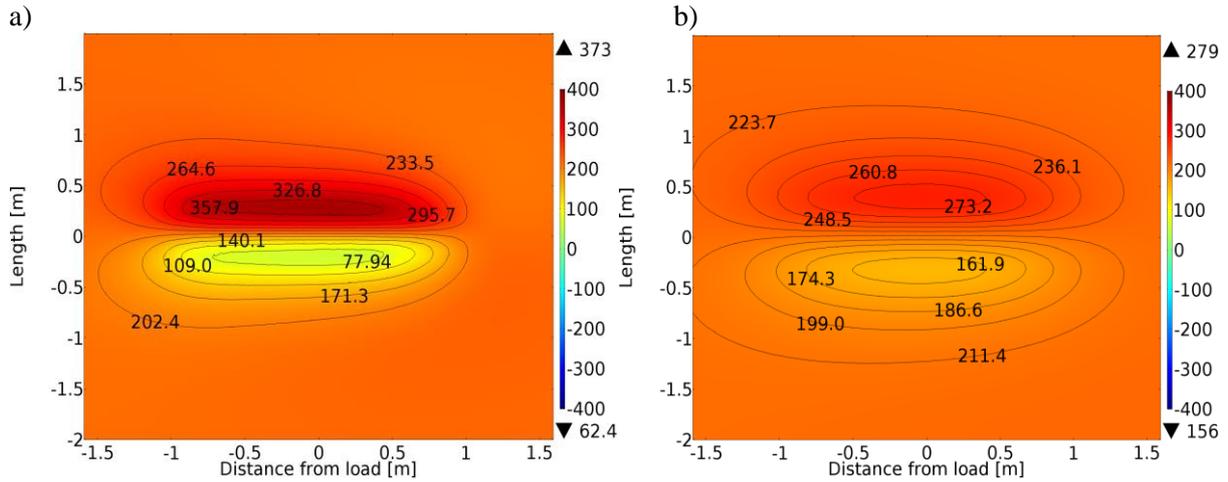

*Figure 13* shear stress with skis down-slope. a) $\tau_{yz}$ Soft-Soft-Soft layer characteristic (F-F-F hand hardness index). b) $\tau_{yz}$ Hard-Soft-Hard (1F-F-1F hand hardness index). Color-code is in Pa and the values at the black triangles are max-min.

The horizontal shear stress $\tau_{yz}$ shown in Figure 14a is focused around the skis but with higher values at the start of the skis on the left in the direction of the slope. The shear stress field is displaced towards the slope. Figure 14b shows four shear stress areas where the two left areas are less focused. All areas are slightly displaced toward the slope.

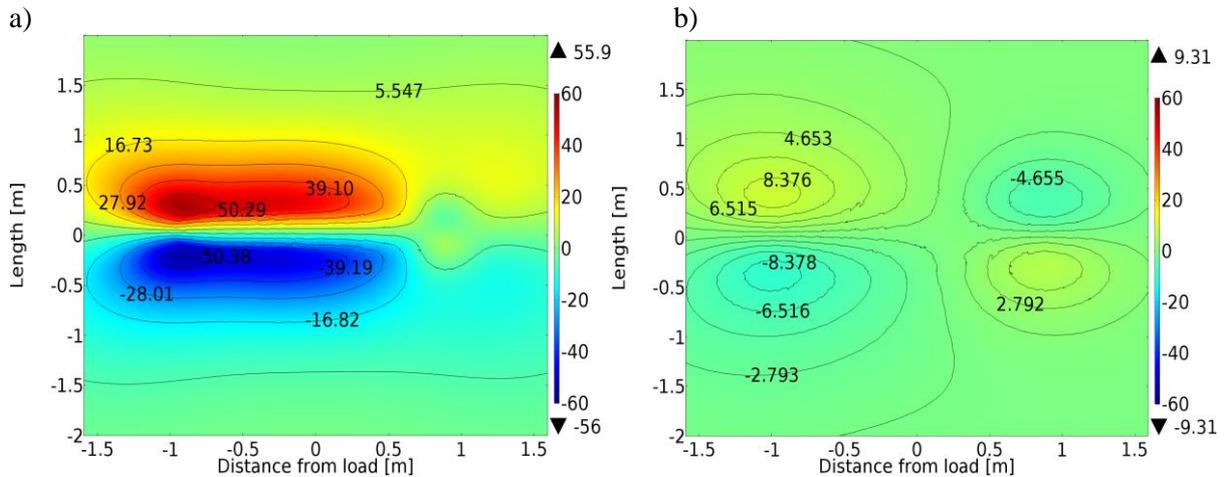

*Figure 14* shear-stress with skis down-slope to the slope without slab weight added a) $\tau_{xy}$ Soft-Soft-Soft layer characteristic (F-F-F hand hardness index). b) $\tau_{xy}$ Hard-Soft-Hard (1F-F-1F hand hardness index). Color-code is in Pa and the values at the black triangles are max-min.

*Weak layer failure criterion*
Considering the volumetric plastic strain as an indication of failure criterion, the area of plastic deformation could be interpreted as a critical crack area to initiate an avalanche.
The following Figure 14 illustrates the volumetric plastic strain on top of the weak-layer. For the 1F-F-1F layer characteristics no plastic strain occurred, and I show only the results of the F-F-F. The plastic strain area in Figure 14a surrounds the skis and is displaced towards the slope. The skis are pointing down-slope. Figure 14b shows the effective plastic strain for the skis parallel to the slope. Only on the left side of the skis has plastic strain occurred with approx. twice the plastic strain value compared to Figure 14a.

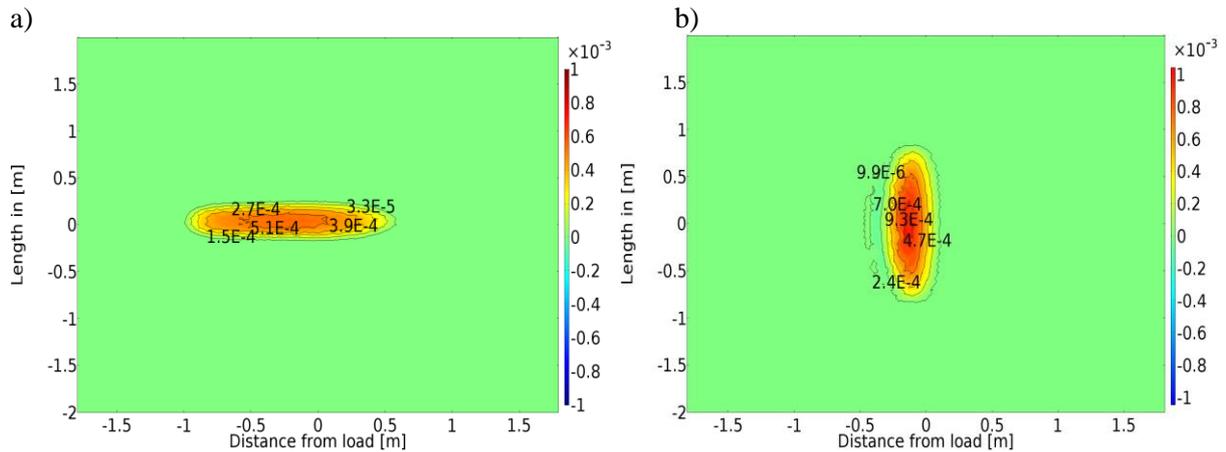
*Figure 15* volumetric plastic strain a) with skis down-slope, b) with skis parallel to the slope.

The last figures in this work are trying to illustrate in which areas the critical crack length is reached. (Gaume and Reuter, 2017) define a critical length $a_c$ using the delta shear stress $\Delta\tau$ and the slope stress $\sigma$ on top of the weak-layer in their 2D models. In Figures 16 and 17 I use the critical crack length equation where each shear-stress value in the weak-layer slice is used to calculate the critical crack length. Figure 16a and b are showing the "crack-length" areas where the skis are pointing down-slope or are parallel to the slope. The "crack-length" area in Figure 16a is around the skis but displaced towards the slope. In Figure 16b the "crack-length" areas are at the start and end of the ski when the skies are parallel to the slope and displaced on the left side of the skis in the direction of the slope.

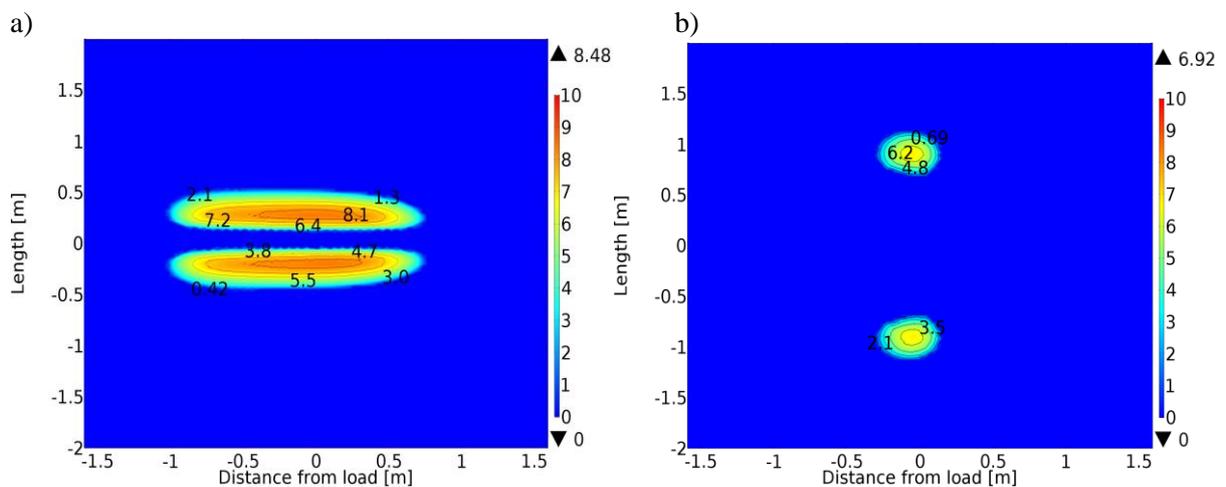
*Figure 16* Crack length $a_c$ calculated from critical shear stress $\tau_{yz}$ downhill (F-F-F), ski parallel to the slope (F-F-F). in cm.

Figure 17 is similar to figure 16 but using the shear stress $\tau_{xz}$. The "critical length" area in Figure 17a surrounds the skis with a clear maximum close to the start of the skis and in the direction of the slope. Figure 17b has a clear focused "critical length" area on the left side of the skis parallel to the slope.

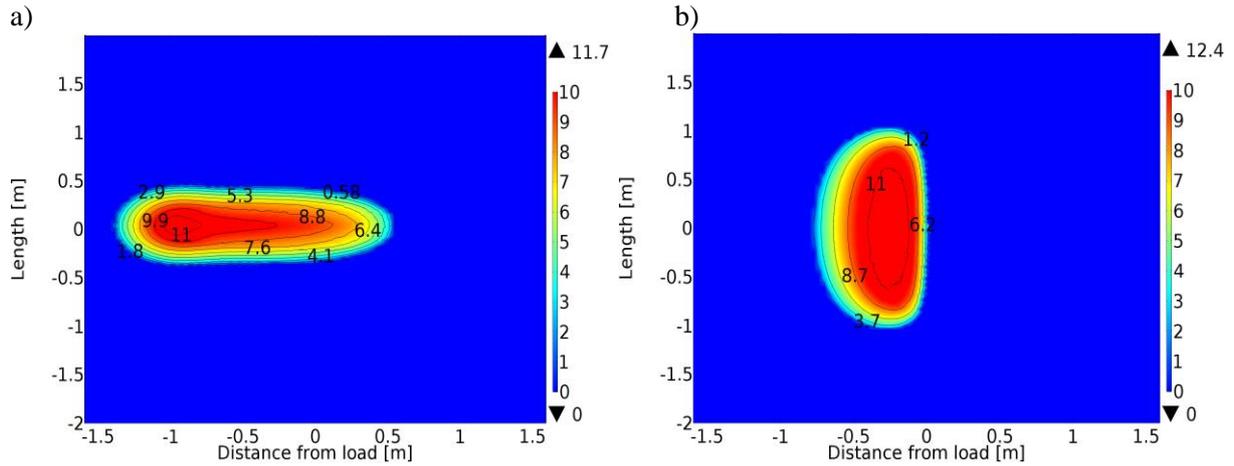

*Figure 17* Crack length $a_c$ calculated from critical shear stress $\tau_{xz}$ downhill (F-F-F), ski parallel to the slope (F-F-F). in cm.

## 5 Discussion

Avalanches triggered by skiers are usually initiated by a fracture in a weak-layer which then propagates along the weak-layer boundary until the slab is released. Furthermore, the specific structure of the layered slab is essential to whether the weak-layer collapses.

The presented 3D simulation results are focused on the weak-layer for two different layer characteristics, where one represents a soft layer structure and the other a hard-soft-hard structure. I used realistic relationships between snowpack properties similar to (Gaume and Reuter, 2017). In this work the slope angle was held constant at 38 degrees, additionally the skier load was held constant to 780N distributed over the two skis. The size of the model volume was selected to overcome boundary effects and to have very small grid space to increase spatial resolution, together with a reduction in simulation time and processing memory. Most of the avalanches were triggered by a weak-layer buried about 0.5 m below the snow surface. Only a few avalanches are triggered by a slab thickness about 1m or more (e.g. van Herwijnen and Jamieson, 2007). The simulation results show no indication of boundary artifacts and the given model volume depth of 1.21 m guarantees a good approximation to real slab layers.

Three different weak-layer properties are simulated in order to better understand the weak-layer failure criterium. The stress field at the weak-layer location was simulated and clearly shows different patterns depending on if the skier is skiing down-slope or traveling parallel to the slope. The shear-stress values are relatively similar between the two skier-directions.

The compressional normal stress field is following the shape of the ski but displaced in the direction of the slope. The Hard-Soft-Hard layer structure unravels and dissipates the stress field along the hard layer and less stress reaches the weak-layer. This is similar to what Schweizer (1993) pointed out, that a hard layer in the snowpack can form a sort of bridge which spreads the induced stress over a larger lateral distance, at the same time decreasing the stress to the layers below the "bridge".

Longitudinal and transversal stress fields below the skier are similar to previous measurements and simulations (e.g. Schweizer and Camponovo, 2001). However, the latter study is only considering the induced stress area below the skier but not showing whether the stress overcomes the weak-layer strength. As shown in the shear stress results the layer-structure of the snowpack has a significant influence on the shear stress propagation in the snow slab. The 3D simulations confirm the bridging nature of the hard snow layer and how the shear stress dissipates and spreads inside this layer (e.g. Figure 8a, b). Below the hard-layer the shear stress propagates over a much more expansive area.

In order to better describe the skier-load induced weak layer collapse, the resulting plastic deformation could be interpreted as a crack area. Snow deformation in the weak-layer occurs primarily in the icy fabric built by the bonds between the grains. Small scale cracks in the snow fabric can cause plastic deformation at low strain rates (e.g. Swinkels, 2017). The space between the grains decreases and the connection between the grains break. This results in an increased stiffness or layer hardening.

The plastic deformation was simulated using elastoplastic behavior of the weak-layer and the volumetric plastic strain results show where the weak-layer was plastically deformed. Similar to (Gaume et al., 2018) I assume that the hardening and softening depends on the volumetric plastic deformation and when the volumetric plastic strain increases the weak-snow layer may fracture under tension. As shown in Figure 15 the volumetric plastic strain is indicating hardening in the weak-layer which will cause a much stiffer weak-layer; and after releasing the induced skier load, a possible fracture.

The critical crack length is a value calculated from the shear- and normal stress (Equation 9) and the characteristic length (Equation 10). Due to the shear-stress the critical crack length occurs in similar areas where the shear-stress is different to the slab weight (Figure 9a and 16c as an example). Considering the length of the volumetric strain area as the skier crack length $l_{sk}$ and using the calculated critical crack length $a_c$, the values of which are presented in figure 16, the stability index can be calculated using equation 11. Following the definition described in (Gaume and Reuter, 2017), with a stability index $S_p < 1$ crack propagation occurs. The simulation results show relatively small critical crack length which lead to a stability index always less than 1. Contrary to the F-F-F layer characteristic the H-F-H layers do not show any volumetric plastic strain deformation and the stress field at the weak-layer is too small to overcome the slab weight, except when the skis are parallel to the slope for the shear stress $\tau_{xz}$ with a maximum value of 2.5cm and ½ of the area size (not shown here).

Considering the plastic deformation as a crack area it is mostly caused by the normal and tensile stress due to compression. The resulting shear stresses were used to calculate the critical crack length and shows a good area agreement with the shear stress area $\tau_{xz}$ but not to $\tau_{yz}$ (Figure 15 compared to Figure 16 and 17). Equation 9 was developed for a 2D slab model where the shear stress in the direction of the down-slope was considered. This critical crack length approach is in good agreement with field measurements (e.g. Gaume et al., 2017, Gaume and Reuter, 2017) and the shear stress in the down-slope direction covers a similar area to the plastic deformation. Considering the area length in the direction of the slope (Figure 15), it can be interpreted that the critical crack length is more dependent on the shear stress down-slope. However, the 3D simulation provides critical crack length for skiing parallel to the slope as well, but the pattern does not match to the pattern or direction of the plastic deformation (volumetric plastic strain).

Reiweger et al. (2015) show a good agreement between the Mohr-Coulomb-Cap model and field data and the resulting plastic deformation shown in my results might be a good approximation to real snow failure behavior.

# 6 Conclusion

In this simulation-based research I tested three different resulting weak-layer properties (shear-stress, volumetric strain and critical crack length) in order to better understand the weak-layer failure criterium. The stress field on top of the weak-layer shows different patterns depending on if the skier is skiing down-slope or traveling parallel to the slope. However, the shear-stress values on top of the weak-layer are relatively similar for both the tested skiing-directions. If the skiing direction has an impact on the risk to initiate an avalanche, the shape of the plastic deformation or the crack-area on top of the weak-layer is a significant factor, as well as the stresses.

Using the Mohr-Coulomb-Cap yield criterium adds plastic deformation into the 3D model. The resulting plastic deformation could be interpreted as a crack area where the weak-layer fabric is brittle compacted. Weak-layer compaction occurs primarily in the icy fabric built by the bonds between the grains. Small scale cracks in the snow fabric can cause plastic deformation at low strain rates which can result in the collapse of the weak-layer by a skier induced load. Considering the plastic deformation as a crack area on top of the weak-layer, the plastic deformation is mostly caused by the normal and tensile stress due to compression. However, the 3D simulation also provides critical crack length for skiing parallel to the slope as well, but the pattern does not match to the pattern or direction of the plastic deformation.

A hard layer in the snowpack forms a type of bridge which spreads the induced stress over a larger lateral distance, at the same time decreasing the stress to the layers below the "bridge". The 3D simulations confirm the bridging nature of the hard snow layer and how the stresses dissipate and spread inside this layer.

The results presented here can be useful for evaluating the stability of skiing down-hill or parallel to the slope, eventually improving avalanche danger estimation and therefore the safety of backcountry skiing pursuits.

# Appendix A

## PARAMETERS

| Name | Expression | Value | Description |
|---|---|---|---|
| def_scale | 10 | 10 | exaggeration |
| D | 0.36 | 0.36 | Slab-thickness |
| R | 780[N] | 780 N | Skier line load |
| phi | 38[deg] | 0.66323 rad | Slope angle |
| ski_z | R*cos(phi) | 614.65 N | Ski load vertical |
| ski_x | R*sin(phi) | 480.22 N | Ski load horizontal |
| l1_E | 0.3[MPa] | 3E5 Pa | Layer 1 Youngs modulus |
| l1_rho | 120[kg/m^3] | 120 kg/m³ | Layer 1 density |
| l2_E | 0.3[MPa] | 3E5 Pa | Layer 2 Youngs modulus |
| l2_rho | 120[kg/m^3] | 120 kg/m³ | Layer 2 density |
| l3_E | 0.3[MPa] | 3E5 Pa | Layer 3 Youngs modulus |
| l3_rho | 120[kg/m^3] | 120 kg/m³ | Layer 3 density |
|  |  |  |  |